\documentstyle[emulapj]{article}

\begin{document}

\title{LEPTON ACCELERATION BY RELATIVISTIC 
COLLISIONLESS MAGNETIC  RECONNECTION}

\author{D.A.Larrabee}

\affil{Physics Department, East 
Stroudsburg University, 
East Stroudsburg, PA 18301-2999; Larrabee@esu.edu} 

\author{R.V.E. Lovelace and M.M. Romanova}

\affil{Department of Astronomy, Cornell
University, Ithaca, NY 14853-6801; RVL1@cornell.edu, 
Romanova@astro.cornell.edu} 

\begin{abstract}

   We have calculated self-consistent
equilibria of a collisionless relativistic 
electron-positron gas in the vicinity of
a magnetic $X-$point.
 For the considered conditions, pertinent
to extra-galactic jets, we find that
leptons are accelerated up to Lorentz factors
$\Gamma_0 = \kappa e B_0L {\cal E}^2  /mc^2 \gg 1$,
where $B_0$ is the typical magnetic 
field strength, ${\cal E}\equiv E_0/B_0$, with
$E_0$ the reconnection electric field,  
$L$ is the length scale of the 
magnetic field, and $\kappa \approx 12$.
   The acceleration is
due to the dominance of the electric
field over the magnetic field
in a  region around the $X-$point.
   The distribution function of the
accelerated leptons is found to be 
approximately $dn/d\gamma \propto \gamma^{-1}$
for $\gamma \lesssim \Gamma_0$.
   The apparent distribution function
may be steeper than $\gamma^{-1}$ due to the distribution
of $\Gamma_0$ values and/or the radiative losses.
  Self-consistent equilibria are found  
only for plasma inflow rates to 
the $X-$point less than a critical value.

\end{abstract}


\section{INTRODUCTION}

The observed radiation of most
large scale extragalactic jets is 
due to  incoherent synchroton
radiation. 
    In a number of  sources,
M87 and 3C 273 for instance, the
radiation lifetime of the electrons
(and possibly positrons)
is much less than the transit time 
from the central source
(Felten 1968). 
   Thus, there must be  mechanisms
for the electron ``reacceleration.''

   Several mechanisms have been
proposed to account for the
reacceleration of the electrons. 
   These include
Fermi mechanism of particle
acceleration (Christiansen, Pacholczyk,
\& Scott 1976), Fermi acceleration in
shock waves (Krimsky 1977;
Axford, Leer, \& Skadron 1978; 
Bell 1977, 1978; Blandford 
\& Ostriker 1978), stochastic 
electric field acceleration (Eilek \& Hughes 1990) 
and whistler wave acceleration (Melrose 1974).
   The effectiveness of the Fermi 
mechanism and stochastic fields 
in accelerating electrons is unknown
(Blandford \& Eichler 1987;
Eilek \& Hughes 1990; Jones \& Ellison 1991). 
    Resistive tearing has been discussed
as a means of producing neutral layers
which can accelerate  electrons 
(K\"onigl \& Choudhuri 1985;
Choudhuri \& K\"onigl 1986).
      Reconnection as a mechanism for
accelerating electrons has been
discussed by a number of authors
(Blandford 1983; Begelman, Blandford, \&
Rees 1984;
Browne 1985; Ferrari 1984, 1985;
Norman 1985; Kirchner 1988; Lesch 1991;
Romanova \& Lovelace 1992). 
   The acceleration of  electrons by
long wavelength electromagnetic
waves trapped in the boundary layer
of a jet was discussed by
Bisnovatyi-Kogan and Lovelace (1995).

   The reconnection of magnetic
fields at a neutral point as
described by magnetohydrodynamics
(MHD) has been studied extensively.
   It has been shown that the
neutral lines  can  evolve into
current sheets (Syrovatskii 1971).
    Resistive MHD simulations of
reconnection by
Biskamp (1986) show a strong tendency
to form a current sheets.

 \begin{figure*}
\centering
\epsscale{1.} 
\figcaption{Geometry of magnetic
field reconnection layer.
} 
\end{figure*}

    Collisionless reconnection
 has been studied by following particle
orbits, assuming nonrelativistic
particles and prescribed fields
(Galeev 1984; Zelenyi et al. 1984;
Burkhart, Drake, \& Chen 1990; Deeg 1991). 
  The particles were launched away
from the $X-$point and
given an initial velocity
that is appropriate for a
given thermal velocity
distribution. 
   Acceleration
of the particles near the $X-$point was observed. 
  This work
was extended to include
the self-consistent fields
of the  particle
flows (Burkhart, Drake, \& Chen 1991).
   For sufficiently high inflow
rates they found an outflow
shock where the  flow
velocity exceeded the Alfv\'en
velocity. 
   In addition they found
that sufficiently large
inflow rates caused the width
of the current layer to
``collapsed to zero,'' 
 and concluded that there was
a maximum infow rate
into the reconnection region. 
     Nonrelativistic particle
trajectories have been
studied in reconnection regions
as part of an investigation
of the Earth's magnetosphere (Speiser 1965).
   This work has been extended
in an effort to define an effective
conductivity in the absence of
particle collisions (Speiser 1970).

   Direct acceleration of particles
in reconnection layers was proposed by
Alfv\'en (1968). 
    Direct relativistic acceleration 
of leptons near the
$X-$line of a magnetic field configuration
was studied by
Romanova  and Lovelace (1992).
   The static electric field 
parallel to this $X-$line can cause
the acceleration of
leptons to the very high Lorentz factors
observed.
   Romanova and Lovelace (1992)
analyzed the particle orbits in the current layer 
structure proposed by Syrovatskii (1971).
    For the case of an electron-positron
plasma they obtained an energy distribution
function $dn/d\gamma \propto \gamma^{-1.5}$.
  Recently, Zenitani and Hoshino (2001) have
done relativistic particle-in-cell simulations
of a reconnection layer in an electron-positron
plasma and have found a spectrum $\gamma^{-1}$
out to $\gamma \sim 20$.

     In this work we solve 
the relativistic
equations of motions for
an electron-positron plasma moving
in the
electric and magnetic
fields of a reconnection
layer indicated in Figure 1.
    In turn, the particle
motion is used to calculate
the current-density and
the associated self magnetic
field. 
  The particle orbits and field
calculations are done iteratively
to give a self-consistent equilibrium
for the reconnection layer.
   Once we have this equilibrium
we can determine
the distribution function of the
accelerated leptons.
    Section 2 of this paper develops
the  theory, and Section 3 discusses the
methods and results.
    Section 4 gives the conclusions of
this work.

\section{THEORY}

\subsection{Physical Picture} 

   The geometry of the stationary
($\partial/\partial t =0$) 
reconnection region
is shown in Figure 1.
The  magnetic field ${\bf B}=B_x(x,y) \hat{\bf x}+
B_y(x,y) \hat{\bf y}$ has 
an $X-$point in the $(x,y)$ plane at $x=0,~y=0$.
    A uniform electric
field exists in the $z-$direction, ${\bf E}=
E_0\hat{\bf z}$ with ${\bf \nabla}\times {\bf E}=0$
and $E_0<0$.
   Therefore,    the lepton gas 
drifts with velocity 
$v_{dy}=cE_zB_x/{\bf B}^2$ in
the $y-$direction towards the
magnetic $X-$point from above and below.
   Because  the gas is electrically neutral
there is no net current due
to this
drift (both electrons
and positrons drift with the same
velocity in the same direction).

   The ${\bf B}$ field vanishes at
the $X-$point, and consequently 
in the vicinity of this  point 
the particle motion is
dominated by the electric field.
   The electric field accelerates  
electrons  in say the $+z$ direction and
positrons in the $-z$ direction.
  This gives a current-density in the $-z$
direction.  
   As the leptons are accelerated in the
$\pm z$ directions their motion is ``bent''
into the $\pm x$ directions by the 
$({\bf v} \times {\bf B}/c)_x$ force.
   At a large enough distance $|x|$
from  the $X-$point the magnetic 
field becomes dominant
and the particles   exhibit the usual
drifts (Northrop 1963).

    The electrons and positrons
are confined near $y=0$ by the $B_x$
magnetic field.
If a particle drifts into the $X$ point
from above, and then overshoots
the $y=0$ plane, the ${\bf E} 
\times {\bf B}$ drift will 
then deaccelerate
the particle and return it
toward the $y=0$ plane. 
   The particles also 
 drift in the
$x-$direction with velocity 
$v_{dx}=-cE_zB_y/{\bf B}^2$.
 Both electrons
and positrons drift away from
the $X-$point. 
   This is due
to the component
of the ${\bf B}$ field in the
$y-$direction.
Since the $ x-$drifts
for electrons and positrons
are in the same direction
there is no net current in
the $x-$direction.
Since both positrons and
electrons enter the $X-$point
from both sides ($y > 0$ and $y < 0$)
the net contribution of the 
$y-$motions cancels, 
and there is no
$y-$direction current.

  We make the simplifying approximation
of neglecting the $y-$thickness of 
the current layer.
    In this case,
the particles move in the $y=0$ plane.
We therefore treat 
equilibria where the particles drift
from above and below into the $y=0$ plane
where they are accelerated in the $\pm z$
directions and then expelled in the 
$\pm x$ directions. 
   Thus we
calculate the particle orbits 
$[x(t),0,z(t)]$.
   Knowing the particle motion
allows us to calculate the surface
current-density ${\cal J}_z(x)$.
   From this we calculate
the self-magnetic field.
    We then use the self
field to recalculate the
particle orbits. 
   We iterate on this process so
that we get a self-consistent
solution of the self magnetic
field and the particle orbits.
   The self-consistent orbits
of the leptons can then be used
to derive the energy spectrum
of the accelerated leptons.

\subsection{Single Particle Motion} 

  The magnetic field can be written as 
$
{\bf B} = {\bf \nabla} \times
[A_z(x,y) \hat{\bf z}],
$
where $A_z$ is the total vector potential.
   For specificity we consider
an $X-$type null point of the 
${\bf B}$ so that $A_z(x,y)$
is an even function of both arguments.
     The magnetic field consists of a
``external'' component due to 
distant currents and the ``self field''
due to local currents. 
   The external component $A_z^{ext}$ is
divergence and curl free. 
  We take the leading terms of a Tayor
expansion of this field, 
$$
A_z^{ext} = {B_0\over 2 L} (x^2-y^2)~,
$$
so that 
$$
B_x^{ext} = B_0 {y\over L}~,\quad
B_y^{ext} = B_0 {x\over L}~,
\eqno(1)
$$ 
 where $B_0 >0$ without loss of generality.
The total field is given by
$$
A^{tot}(x,y) = A^{ext}+A^{self}~.
\eqno(2)
$$
$A^{tot}$ is  an even function
of both arguments but its dependence on
$(x,y)$ is changed by the self-field.

   We consider quasi-stationary conditions
so that
${\bf \nabla} \times
{ \bf E} = 0$ and thus
$
{ \bf E} = -{\bf \nabla} \Phi,
$
which is the ``external'' electric field.
  The relevant solution  is
${\bf E} =  E_0 \hat {\bf z}$
and $\Phi=-E_0 z$, 
where $E_0={\rm const} <0$.  
   This corresponds to 
plasma flowing into the $X$ point
from above and below the $y=0$ plane.
  Because we consider an electron-positron
plasma the ``self electric field'' is 
zero, and the total electric field is $E_0\hat{\bf z}$.

  The single particle motion is
described by the Lagrangian
$$
{\cal L} = -mc^2 \sqrt{1- {\beta}^2}+
{q \over c}{{ A_z}  { v_z}} - q\Phi~,
\eqno(3)
$$ 
where $\beta \equiv |{\bf v}| / c$ 
and the electrostatic potential $\Phi = -E_0 z$. 
The corresponding Hamiltonian is
$$
{\cal H }=\bigg[({\bf P}-{q\over c}
{\bf A})^2+m^2c^4\bigg]^{1\over2}+q\Phi~,
$$ 
where ${\bf P}$ is 
the canonical momentum. 
 Because $\partial {\cal L}/\partial t=0$,
the ${\cal H}=$ const which is the single
particle energy.
 
  A further  constant of the motion is the
canonical momentum 
in the $z-$direction, $P_0$.
   Because $A_z$ is independent of
$z$,
$$
{d P_z \over dt} = 
-{\partial {\cal H} \over \partial z} = qE_0~,
\quad {\rm or} \quad {d \over dt}(P_z-qE_0t) =0~,
$$ 
so that
$$
P_0 \equiv P_z-qE_0t= m\gamma {dz \over dt}+
{q \over c}A_z-qE_0t ={\rm const}~. 
\eqno(4)
$$
The $z-$equation of motion is simply
$d P_0/dt = 0$ which gives
$${d \over dt} 
\left(m\gamma {dz \over dt}\right) 
+{q \over c}
{B_y}{dx\over dt}-{q \over c}
{B_x}{dy\over dt}-qE_0 =0~, 
\eqno(5)
$$
where $\gamma =(1-\beta^2)^{-1/2}$.
   For the limit where the
$z-$acceleration is small,
equation (5) gives
$$
{q \over c}B_y{dx \over dt} -
{q \over c}B_x {dy\over dt}- qE_0 = 0,
$$
or
$$
{\bf v} = c {\bf E} \times {\bf B}/{\bf B}^2
\eqno(6)
$$
which is the well-known ${\bf E}\times {\bf B}$
drift velocity (Northrop 1963).

 We assume
that away from the $X-$point the
leptons are relatively `cold' 
in the sense that their thermal
velocity spread is less than 
their ${\bf E \times B}$ drift
velocity  which is less than $c$.
   Thus, leptons which are initially
above (below) the $(x,z)$ plane 
drift down (up) towards this 
plane as sketched in Figure 1.  
    A restricted class of solutions 
has the form ${\bf x}(t) =
[x(t),0,z(t)]$. 
   A particle
``launched'' from a position
$(x_0,0,0)$ in the $y=0$ plane
with $v_y=0$
stays in this plane.
   Of course, a particle  launched away from
this plane (or one that has $v_y\neq 0$),
 oscillates about
the $y=0$ plane.

   In this work we neglect 
the thickness of the current layer in
the $y$ direction. 
  Therefore, we use the particle orbits 
in the $y=0$ plane.
 We consider a set of particles
entering the $y=0$ plane from above
and below
with initial locations given
by $(x_0,0,0)$.

     Dimensionless variables are introduced as
$$
\hat x \equiv {x / L}~,
\quad \hat t \equiv \omega_0 t~,\quad
{\rm where}~~ \omega_0 \equiv {|q| B_0 / mc}~,
\eqno(7)
$$
is the non-relativistic cyclotron frequency in
the reference field $B_0$.
   Dimensionless electric and magnetic fields
are naturally measured in units of $B_0$.  
   In particular,  we let
$$
{\cal E} \equiv |E_0|/B_0~.
\eqno(8)
$$    
  A characteristic Lorentz factor can be
defined as
$$
\gamma_0 \equiv {|q| B_0 L \over m c^2} 
={\omega_0 L \over c}~.
\eqno(9)
$$
  We can rewrite this as
$$
\gamma_0 \approx 5.9\times 10^3 
\left({B_0 \over 10^{-6}{\rm G}}\right)
\left( {L \over 10^{13} {\rm cm}} \right)~.
$$
Note that  $v_x/c=\gamma_0 (d\hat x/d \hat t)$, for example.
We let $n_\infty$ be the number density of leptons
(electrons $+$ positrons) at a large distance
from the $X-$point.  
  An important dimensionless
quantity is
$$
\zeta \equiv {4\pi n_\infty \gamma_0 m c^2  \over B_0^2}~.
\eqno(10)
$$
The role of this parameter is described in \S 3.

\subsection{High Energy Approximation}

   A particle launched at $x_0$ in the $y=0$ plane
is accelerated in the $z-$direction
 by the $E_0 \hat{\bf z}$ field
and it drifts in the $x-$direction because of
the ${\bf v} \times {\bf B}$ force.
   The closer to the $X-$point a
 particle starts  (the smaller $x_0$ is) 
the higher the energy it is
accelerated to. 
   An approximate 
solution for the particle 
motion in the external  magnetic
field ${\bf B}^{ext}$ is possible for 
sufficiently small $x_0$
where there is a large acceleration in 
the  $z-$direction.

   Assuming that the particle
is already highly  relativistic we let
$ 
p_x(t) =m  \gamma c \sin(\theta)$ and
 $p_z(t) = m \gamma c \cos(\theta)$.
   For specificity we consider
the acceleration of electrons.
The equations of motion become
$$
{{dp_x} \over dt}=-qB_y^{ext}(x) \cos(\theta )~,
~~
{{dp_z} \over dt}=qB_y^{ext}(x)
\sin(\theta ) +q E_0~.
$$ 
We assume $\theta^2 \ll 1$ and discuss
the conditions for this later. 
The energy equation  gives
$m\gamma c^2 =qE_0 z +{\rm const}$
and the small angle approximate gives
$t\approx z/c$.

   The $x-$equation of motion in dimensionless
variables   becomes
$$
{{d}\over d\hat z}\left( \hat z{{d \hat x} \over d\hat z}
\right)= {\hat x \over {\cal E}}~.
\eqno(11)
$$ 
The relevant solution to equation (11) is
$$
\hat x=\hat x_0 I_0
\left(2\sqrt{{\hat z/ {\cal E}}}\right)~,
\eqno(12)
$$
where $I_0$ is the usual modified Bessel function.
   This dependence is valid for small angles, that is,
for $(d\hat x /d \hat z)^2 \ll 1,$ or 
$[\hat x_0I_1(2\sqrt{\hat z/{\cal E}})
/\sqrt{\hat z {\cal E}}]^2
\ll 1$.   
    A necessary condition for this is that
$\hat x_0  \ll {\cal E}$.  
    The geometry of the orbits for highly
relativistic motion depends  only on $\hat x_0$ and
${\cal E}$ (in fact, on
just ${\cal E}/\hat x_0$) and not for example on $\gamma_0$.

Figure 2 shows  orbits  in the external
magnetic field calculated without approximation.
  The approximation of equation (12) is also shown.

 \begin{figure*}
 \caption{Sample electron orbits in the $y=0$ plane
for the external magnetic field $B_y^{ext}=B_0~x/L$
for the case ${\cal E}=0.5$, $\gamma_0=10^3$,
and $\hat x_0 =0.1,~0.2,$ and $0.3$.
   For large $x$, where
the orbits are ``looping,'' the drift in the $+x-$direction
is the $E_z \times B_y$ drift, and the drift
in the $+z$ direction is the gradient $B$ drift
$\propto B_y \times {\bf \nabla} B_y(x,0)$.
The dashed line segment is the approximate dependence
of equation (12).  
  The positron orbits are mirror
images for $z \rightarrow -z$.
} 
\end{figure*}

\subsection{Current-Density 
from  Single Particle Motion} 

Far  from the $X-$point, ideal
magnetohydrodynamics applies, 
and the fluid particles are confined 
to flux surfaces where $A_z=$const. 
   Even though this is a time 
independent situation, the flux
surfaces drift inward 
towards the neutral layer with a 
velocity equal to ${\bf v}_d=c{\bf E} 
\times {\bf B}  / B^2 $. 
  Since the $\bf B$ field is tangent to 
surfaces of constant $A_z$ and 
the electric field is in the 
$\hat{\bf z} $ direction, the field 
lines drift inward towards the 
$X-$point at the same speed the 
particles do.  
   Near the $X$ point, where
$|{\bf E}|> |{\bf B}|$, ideal magnetohydrodynamics
breaks down. 

  We assume that after a particle
inflows (from above or below) to the $y=0$ plane it gets
launched on an orbit in this plane, $[x(t),0,z(t)]$.
   The initial position of the particle
in the $y=0$ plane is  $x_0$ and some value of $z$.
    We denote the number flux density of particles
(electrons $+$ positrons) 
inflowing from above and below the $y=0$ plane 
as ${\cal F}(x_0)$ (number per ${\rm cm}^2$ per second). 
   The simplest case is that of a uniform density of
inflowing plasma where 
$$
{\cal F}=2 n_\infty c {\cal E}~,
\eqno(13)
$$
where $n_\infty$ is the number density ($1/$cm$^{3}$)
of electron plus positron
density at large distances and where the factor
of two comes from the two sides of the current layer. 
  The motion of a single particle 
starting from position $x_0$ is described
 by its position $[x(t|x_0), 0, z(t|x_0)]$ and  its
velocity $[v_x(t|x_0), 0, v_z(t|x_0)]$.

     The surface current-density (charge per cm 
per second) for each initial $x_0$
is then given by 
$$
{\cal J}_z(x|x_0)dx_0 = 
q {\cal N}(x|x_0) \langle v_z(x|x_0) \rangle dx_0~,
\eqno(14)
$$ 
where ${\cal N}(x|x_0)$ (with units $1/$cm$^{3}$)
 is the surface number density
(of  electrons
and positrons) at $x$ 
launched between $x_0$ and $x_0 + dx_0$. 
  In equation (14), $v_z$ is the electron velocity,
$q$ is the electron charge, and the angular brackets
indicate a time average.
   In the considered stationary 
state which is uniform in $z$, 
all averages are independent of $t$ and $z$. 

   Consider particles 
moving in the $+x$ direction (the $-x$ direction
case follows by symmetry), 
then conservation of particles
implies that 
$$
{\cal N}(x|x_0) \langle v_x(x|x_0)\rangle= 
{\rm const} = {\cal F}(x_0)~,
\eqno(15)
$$
where $v_x(x|x_0)$ is the velocity of an electron
{\it or} a positron. 
  Thus, 
$$
{\cal J}_z(x) = \int_0^{x_0^m}dx_0 ~{\cal J}_z(x|x_0)~,
\quad\quad \quad \quad
$$
$$
\quad \quad \quad = q \int_0^{x_0^m}dx_0~ {\cal F}(x_0)~
{\langle v_z(x|x_0)\rangle\over 
\langle v_x(x|x_0)\rangle  }~,
\eqno(16)
$$ 
where $x_0^m$ is the maximum of $x_0$ assumed to
be ${\cal O}(L)$.
The total current between 
$x$ and $x+\delta x$ can be written as 
$$
\delta x~{\cal J}_z(x)  = q \int_0^{x_0^m}dx_0 ~{\cal F}(x_0)
\int_{x}^{x+\delta x} dx~ 
{\langle v_z \rangle \over \langle v_x \rangle}~. 
$$ 

For a particular value of $x_0$ 
a particle orbit passes 
between $x$ and $x+\delta x$ for time between 
 $t_1=t(x|x_0)$ and $t_2=t(x+\delta x|x_0)$ 
(and possibly for time between $t_3$ and $t_4$ and
$t_5$ and $t_6$, or an odd number of intervals).
    This allows us to replace the 
spatial integration by a 
temporal integration over the particle orbit.
Notice that $\langle v_x \rangle =\sum \delta t_j v_{xj}
/\sum \delta t_j$, where $j=1$ or $j=1,2,3$, etc.
represents the different traversals of the interval
$\delta x$ at the distance $x$, and where $\delta t_j =
\delta x/|v_{xj}|$.  We have for example $\int dt v_x=
\sum \delta t_j v_{xj} =\delta x$.
  Therefore, we find
$$
\delta x~ {\cal J}_z(x) = q \int_0^{x_0^m}dx_0 
~{\cal F}(x_0) 
\int_{t_1}^{t_2} dt~ v_z[x(t),z(t)|x_0] ~.
\eqno(16a)
$$ 
The time integral 
needs to be done for all values of $t$
for which the particle is between
$x$ and $x+\delta x$.
    We can put equation (16a) into
dimensionless form using equations (7) - (10)
and  $\hat{\cal J}_z \equiv {\cal J}_z/(cB_0)$.
   For the case where equation (13)
applies this gives
$$
\delta \hat x ~\hat{\cal J}_z  =
\bar \zeta\int d \hat x_0  \int d\hat t ~ 
{d \hat z \over d \hat t}~.
\eqno(16b)
$$
Here, 
$$
\bar \zeta \equiv {{\cal E}\zeta \over 2\pi}~,
\eqno(17)
$$ 
with $\zeta$ 
defined by equation (10), is a 
dimensionless measure of
the rate of inflow of plasma to the
neutral layer.

\subsection{Self-Magnetic Field} 

    For the considered  thin current
layer with surface current-density ${\cal J}(x)$,
we can express the self-magnetic field of this
current as
$$ 
B_y^{sel\!f}(x) = {2 \over c}~{\cal P}\!\! 
{ \int dx^\prime~ {{\cal J}_z(x') \over x-x'} }~,
\eqno(18)
$$ 
where ${\cal P}$ is the principle value.
   In dimensionless variables this becomes
$$
\hat B_y^{sel\!f}(x) = 2~ {\cal P}\!\! 
{ \int d \hat x^\prime~ {\hat{\cal J}_z(\hat x') 
\over \hat x- \hat x'} }~,
$$
The total magnetic field is
$$
B_y^{tot}(x)=B_y^{ext}(x)+B_y^{sel\!f}(x)~.
\eqno(19)
$$ 
The dimensionless form of this equation
is the same but with hats over the
magnetic fields.  

   In the following  we simplify the
notation by dropping the hats and dropping
the $y$ and $z$ subscripts on $B_y$ and ${\cal J}_z$.

\section{METHODS AND RESULTS}

  The self-consistent magnetic field of 
the current layer is calculated iteratively.
   For a relatively weak self field - 
sufficiently small
$\bar \zeta$ compared with unity - a direct
iteration scheme has been found to converge.  
This scheme may be described as
$$
B^{tot}_n ~~ \underbrace{  { \longrightarrow} }_{\rm orbits}
~~ {\cal J}_{n+1}~~ \underbrace{\longrightarrow}_{\rm HT}~~
B_{n+1}^{tot}~~ 
\underbrace{ {\longrightarrow} }_{\rm orbits}~~{\cal
J}_{n+2}~~...~, 
\eqno(20)
$$
where 
$$
B^{tot}_n = x+ B^{ sel\!f}_n~,\quad B^{sel\!f}_n \equiv
2~{\cal P}\!\!\! \int dx^\prime ~
{{\cal J}_n(x^\prime) \over x -x^\prime}~.
\eqno(21a)
$$
Here,   ``HT'' stands for the Hilbert  transform,
and ``orbits'' indicates the evaluation
of equation (16b) using an $x-$grid of $100-200$
points and the calculation of $100-200$
electron orbits.  A stretched $x-$grid is used
with higher resolution at small $x$.  
   The distribution of inflowing plasma ${\cal F}(x_0)$
is assumed uniform in $x_0$ as given by 
equation (13).
   All cases considered here have $\gamma_0 \gg 1$
so that the initial, non-relativistic
or modestly relativistic 
conditions ($\gamma < 2$) of  particles
at their starting points $x_0$ do not affect
the reported results.
   In the iteration sequence we have found
it advantageous to make analytical fits
to the self fields $B_n^{sel\!f}$ which are
used in the subsequent orbit calculations.   

  For larger self-fields we use the iteration (20)
with equation (21a) replaced by
$$
B^{tot}_n = x+ \alpha B^{ sel\!f}_n+
(1-\alpha)B_{n-1}^{sel\!f}~,
\eqno(21b)
$$
where $0<\alpha<1$ is a numerical factor,
and $B^{sel\!f}_n $ is still given by equation (21a).
   This iteration scheme is known to allow
the calculation of self-consistent particle
ring equilibria with large self-fields
(see for example Larrabee et al. 1979, 1982).

The key dimensionless parameters which
determine the self-consistent current layers are
$$
{\cal E}~,\quad \bar \zeta~, \quad {\rm and } \quad \gamma_0~.
\eqno(22)
$$
For the considered limit
$\gamma_0 \gg 1$ and fixed ${\cal E}$, we find that the 
current density and self-field are almost
independent of $\gamma_0$.
   However, the distribution function of  accelerated
leptons does depend on $\gamma_0$ 
and ${\cal E}$ as discussed below.

\subsection{Weak Self-Field}

 \begin{figure*}
\caption{Iteration sequence of the dimensionless
current density for a case of weak self-field
where $\bar{\zeta}=0.05$.   For this case ${\cal E}=
0.5$ and $\gamma_0=10^3$.
} 
\end{figure*}

   Figure 3 shows the current density profiles ${\cal J}(x)$
for a weak field case $\bar \zeta =0.05$ where the convergence
of the direct iteration, equations (20) and (21a), 
is quite rapid.
    The main peak of the current density is 
due to the  direct acceleration of electrons
and positrons in the region 
where $|{\bf E}|/|{\bf B}| >1.$
    The ``shoulder'' of the current density
profile is due to the gradient $B$ drift in
the $z-$direction.
  Figure 4 shows the final profiles of the total
and self-magnetic fields.  
   The self-field acts to reduce the slope of
the total field $B^{tot}$ near the origin.  
    For vanishing self-field the slope is unity.
For the case of Figure 4, the slope is
$dB^{tot}/dx|_0 \approx 0.680$.

   The energy distribution or spectrum  of the leptons 
accelerated in the current layer 
$d n/d\gamma$ is obtained from 
the calculation of the particle orbits.
  The  particles inflowing to the current
layer are uniformly
distributed in $x_0$ and initially
they are assumed to have $\gamma < 2$.
   The Lorentz factor of a particle
as it exits the current layer 
(say, $x >x_{max}$) is therefore
a function of $x_0$,
$\gamma_m = \gamma_m(x_0)$.
   As part of our
orbit calculations we determine
$\gamma_m(x_0)$.
  We find that $\gamma_m(x_0)$ is
a monotonically decreasing function
of $x_0$.
   With $dn/dx_0=$ const  the number of 
particles launched between 
$x_0$ and $x_0+dx_0$, 
we have
$$
{dn \over d\gamma}={dn/dx_0 \over \big| d\gamma/dx_0\big|}~,
\eqno(23)
$$ 
for $\gamma>2$, where we have dropped the $m-$subscript
on $\gamma$.
     
    Figure 5 shows  lepton distribution for 
the same case as Figures 3 and 4. 
  As mentioned earlier the magnetic field of the current
layer is almost independent of $\gamma_0$.
   For these values of $\bar \zeta$ and ${\cal E}$,
we have run a range of  values of
$\gamma_0 = 10^2 -10^4$, and find that the 
distributions
are roughly fitted by 
$$
{dn \over d\gamma } \approx 
{K \over \gamma}\exp\left(-{\gamma \over \Gamma_0}\right)~,
\eqno(24)
$$
where $\Gamma_0=\kappa ~{\cal E}^2~ \gamma_0$ and
$\kappa \approx 12$.
    With $K\approx 1/\ln(0.56\Gamma_0)$ 
the distribution
is normalized to unity.
   The distribution (24) is {\it not} expected to
be the same as  that observed because of
the influence of 
the distribution of $\gamma_0$ values and/or
the affect of radiative losses.
   These affects on the spectrum are 
discussed in \S 3.4.

 \begin{figure*}
\caption{Dimensionless self-consistent 
magnetic field profiles at the fourth iteration 
for same case as Figure 3, $\zeta=0.05$, ${\cal E}=0.5$,
and $\gamma_0=10^3$.  The dashed 
$B^{sel\!f}\bar \zeta$ curve 
 is the analytic fit given
by $5.4x(1+1.9x^2)/(1+0.52x^2+2.1x^4)$.
Near the origin the slope
of the total field, $dB^{tot}/dx|_0 \approx 0.680$,
is reduced from unity owing to the self-field
of the current layer.
} 
\end{figure*}

\subsection{Strong Self-Field}

   For larger self-field strengths $\bar \zeta$, we use
the iteration indicated by 
equations (20) and (21b) usually with $\alpha=0.5$.
  For $\bar \zeta=0.1$ we have not succeeded in getting
the iterations to converge.  
   We find that the
iterations lead to negative values of $B^{tot}(x)$
for $x>0$ which would cause trapping of particles in 
the $x-$direction.
   Such a configuration is inconsistent with
our assumption that the accelerated particles are expelled
in the $x-$direction.
    The existence of a maximum of $\bar \zeta$,
which measures the rate of plasma inflow to
the neutral layer, is consistent with the finding
of a maximum plasma inflow rate by
Burkhart et al. (1991). 
     For $\bar \zeta=0.075$, equations (20) and (21b)
converge after five iterations.
   We find that the maximum of the current-density
${\cal J}(0)$ and the half-width at half-maximum
of the current-density $\Delta x$ increases 
as $\bar \zeta$ increases.  For this case
we find $dB^{tot}/dx \approx 0.412$.
   The energy spectrum of the accelerated leptons
is similar to equation (24).

\subsection{Scalings}

  In more detail, we find the scaling relations
$$
{\cal J}(0) \approx 1.53 ~\bar \zeta~ {{\cal E} / B^\prime}~,
\eqno(25a)
$$ 
and
$$
  \Delta x 
\approx  1.48{\cal E}/\sqrt{ B^\prime}~,
\eqno(25b)
$$ 
where $B^\prime \equiv dB^{tot}/dx|_{x=0}$.
   The first relation can be derived from the
analytic orbits of \S 2.3.  
    For the external magnetic field 
$B^{tot}=x$, we have from \S 2.4,
$$
\lim_{x \rightarrow 0} {\cal J}(x) = \lim \bar \zeta
\int_0^x {dx_0 \over dx/dz}=
\lim \bar \zeta \int_0^x
{dx_0~ \sqrt{z {\cal E}} \over x_0 I_1(2\sqrt{z/{\cal E}})}
$$
$$
=\bar \zeta {\cal E} \int_0^\infty {dz' \over I_0(2\sqrt{z'})}
\approx 1.53 \bar \zeta {\cal E}~,
\eqno(26)
$$
where the relation between $x,~x_0$, and $z$ is given
by equation (12).  
   In the magnetic field $B^{tot}=B^\prime x$, equation
(26) is modified by the replacement ${\cal E}
\rightarrow {\cal E}/B^\prime$.  
  Equation (25b) is
empirical.

    We can use equations (25a)  and (25b)
to obtain an approximate
constitutive equation for the current layer.
Note that $dB^{tot}/dx|_0 = 1 +dB^{sel\!f}/dx|_0$,
and that $dB^{sel\!f}/dx|_0 =-2 
\int dx [{\cal J}(0) - {\cal J}(x)]/x^2$.
Notice in turn that this integral is proportional
to ${\cal J}(0)/\Delta x$.
From this we obtain the  relation
$$
\bar \zeta \approx 0.205
\left[(B')^{1/2} - (B^\prime)^{3/2} \right]~.
\eqno(27)
$$
This dependence is shown in Figure 6 along
with the calculated equilibria.  
    The left-hand part of the curve 
is dashed because we have not found the
corresponding self-consistent equilibria.
    The maximum of $\bar \zeta$ is
thus $\bar \zeta_{max} \approx 0.08$.
  This maximum represents a maximum inflow
rate to the reconnection region analogous 
to that found by Burkhart et al. (1991).

 \begin{figure*}
\caption{Spectrum of leptons accelerated in
the current layer for $\bar \zeta =0.05$, ${\cal E}=
0.05$, and $\gamma_0=10^3$ and $3\times 10^3$.
   For this spectrum we have taken $x_{max} =2.5$.
The dotted curve is an approximate fit for
the $\gamma_0=10^3$ case given by $dn/d\gamma
=(K/\gamma)\exp(-\kappa \gamma/\gamma_0)$
with $\kappa=1/3$.  
   With $K\approx 1/\ln(0.561\gamma_0/\kappa)$, the lepton
distribution function is normalized to unity.
} 
\end{figure*}

 \begin{figure*}
\caption{Constitutive relation for relativistic
electron-positron reconnection from equation (27).
The solid circles are calculated equilibria for
${\cal E}=0.5$ while the open circles 
are for ${\cal E}=0.2$.  
  No equilibria have been found along the dashed
part of the curve.
} 
\end{figure*}

\subsection{Apparent Spectra}

   The observed lepton distribution 
$f(\gamma)$  deduced
from the  synchrotron spectrum of a radio source
will in general be different from the distribution
of  accelerated particles at a given reconnection 
site ($dn/d\gamma$).  For a typical radio source
$f \propto \gamma^{-2.5}$.
   One effect is due to the distribution of $\Gamma_0$
values owing mainly to the distribution of $L$ for
different reconnection sites.
   Note that $k=2\pi/L$ is the wavenumber of
the spatial power spectrum of the ${\bf B}$ field.
With  the distribution $W(\Gamma_0)$, the average
lepton distribution function $f(\gamma)$ for
many reconnection sites is
$$
f(\gamma)=\int d\Gamma_0 W(\Gamma_0) 
{dn (\gamma|\Gamma_0)\over d \gamma}~.
\eqno(28)
$$
For $W(\Gamma_0) \propto \Gamma_0^{-q}$, 
equation (28) gives $f \propto \gamma^{-1-q}$
for $q>1$, where the logarithmic dependence
of the normalization
factor $K$ of $dn/d\gamma$ is neglected.

  The observed $f(\gamma)$ may also be steeper than
$dn/d\gamma$ owing to synchrotron and/or inverse
Compton radiation (see, e.g., Pacholczyk 1970).
   In a steady state the radiative loss
term in the Boltzmann equation for the  leptons,
$\partial [ \dot{\gamma} f(\gamma)]/\partial \gamma$ 
balances the source term $dn/d\gamma$ on the right
hand side, where $\dot{\gamma} \propto -\gamma^2$
describes the radiative losses.
   Thus
$$
f(\gamma) = {{\rm const} \over \gamma^2} 
\int_\gamma ^\infty d\gamma^\prime 
{dn \over d\gamma^\prime}~.
\eqno(29)
$$
This integral is easily done for the case
where $dn/d\gamma$ is given by equation (24).
One finds that the $f(\gamma)$ is
somewhat steeper than $\gamma^{-2}$
for $\gamma < \Gamma_0$ and rapidly
decreasing for $\gamma>\Gamma_0$.

\section{CONCLUSIONS}

   We have calculated the self-consistent 
current-density distribution in the 
region of collisionless 
reconnection at an $X-$type magnetic
neutral point in a relativistic
electron-positron plasma.
    The energy 
distribution function of the
accelerated leptons has also been
calculated.

   We find that there are three 
important dimensionless parameters:
 (1) The characteristic Lorentz factor
$\gamma_0 \equiv e B_0 L/m_ec^2 \gg 1$,
where $B_0$ is a reference magnetic field strength
and $L$ is the length scale of the  magnetic field;
(2) The ratio of the static electric field $E_0$
to the reference magnetic field, ${\cal E} \equiv
|E_0|/B_0$;  and
(3) the strength parameter $\bar \zeta \equiv 
({\cal E}/2\pi)4\pi n_\infty m c^2/B_0^2$,
which is a measure of the rate of inflow of
the plasma towards the reconnection region.
   We find that the current-density distribution
depends on ${\cal E}$ and $\bar \zeta$ but
only weakly on $\gamma_0$.
   The distribution function of the accelerated
leptons is found to be 
approximately $dn/d\gamma \propto \gamma^{-1}$
for $\gamma \lesssim \Gamma_0=\kappa {\cal E}^2\gamma_0$
where $\kappa \approx 12$.
  The simulations of Zenitani 
and Hoshino (2001) give a distribution $\propto \gamma^{-1}$
for $\gamma <20$.
   The apparent distribution function
may be steeper due to the distribution
of $\Gamma_0$ values and/or the radiative losses.
   Self-consistent equilibria are found  
only for plasma inflow rates $\bar \zeta$ to 
the $X-$point less than a critical value
which depends on ${\cal E}$.

\acknowledgments{
This work was supported in part by NASA grants
NAG5-9047 and NAG5-9735 and by NSF grant AST-9986936.
}

\end{document}